\documentclass[prd,reprint,prd,twocolumn,superscriptaddress,nofootinbib]{revtex4-2} 
\usepackage{graphicx,amsfonts,amsmath,amssymb,amsthm}% Include figure files
\usepackage{dcolumn}% Align table columns on decimal point
\usepackage{bm}% bold math
\usepackage{subcaption}
\begin{document}

%\preprint{APS/123-QED}

\title{New topological phase transitions in $\kappa$-deformed $\phi^4$ lattice models}% Force line breaks with \\

\author{Adith A}
 \affiliation{Department of Physics, National institute of technology Calicut, Calicut, 673601, India}%Lines break automatically or can be forced with \\
 \email{adithammath@gmail.com}
\author{A. Bhagwat}
 \email{ameeya@cbs.ac.in}
\affiliation{
School of Physical Sciences, UM-DAE Centre for Excellence in Basic Sciences, University of Mumbai, Vidyanagari campus, Mumbai 400098, India
}
\affiliation{Centre for Excellence in Theoretical and Computational Sciences, University of Mumbai, Vidyanagari Campus, Mumbai 400098, India}

\date{\today}% It is always \today, today,

\begin{abstract}
This paper explores topological phase transitions in the momentum space of the $\kappa$-deformed $\phi^4$ lattice model within the mean-field approximation. We show that the double-well structure of the momentum space gives rise to degenerate momentum states at high energies and that this change is marked by a phase transition. We investigate this transition using both numerical and analytical methods, with the analytical approach based primarily on singularity theory. Our numerical analysis shows that the energy at which the topological phase transition occurs is related to the noncommutative parameter by the simple power law $E \sim 1/\alpha^{4}$. By mapping the constant-energy-level surface to the $A_{3}$ singularity, we obtain the same result analytically. We also present results unique to noncommutative systems, showing that the fraction of particles in degenerate momentum states has a nonzero commutative limit. As a nonperturbative effect, this result provides a testable prediction for the existence of noncommutative spacetime.
\end{abstract}

\maketitle
\section{Introduction}\label{sec1}
The last few decades have seen an increasing interest in noncommutative geometry and its various physical realizations. It was first suggested in 1930 by Heisenberg \cite{PauliBriefwechselVol2,PauliBriefwechselVol3} to solve UV and IR divergences in various quantum field theories, and was formalized by Snyder \cite{Synder,JACKIW200230} in 1947. However, this idea was replaced by the then popular `renormalization schemes'. Interest in noncommutative spaces was revived after it was realized that quantum field theory on noncommutative spacetime emerges as a low energy limit of the theory of open strings \cite{NathanSeiberg_1999}. Another motivation to study noncommutative spacetime is when one considers the Planck scale structure of spacetime. At this scale the idea of spacetime as a smooth manifold fails, and adding noncommutativity gives a way to establish a fundamental length scale.

The theory has its foundations in noncommutative geometry introduced by Alain Connes in his foundational work \cite{connes1995noncommutative} on the subject where he used noncommutative algebras to study such spaces using techniques from $C^{*}$-Algebra, Von Neumann algebra etc. At the heart of the theory lies the noncommutative extension of the Gelfand-Naimark theorem which gives us a correspondence between noncommutative spaces and noncommutative $C^{*}$ algebras. The interested reader is referred to mathematical books on the subject \cite{connes1995noncommutative,Connes2,khalkhali}.

The term `noncommutative geometry' is a broad one and is the general term for various models of noncommutative spaces like Moyal spacetime \cite{RevModPhys.73.977,NathanSeiberg_1999}, fuzzy spheres \cite{fuzzysphere}, Quantum planes \cite{WESS1991302}, $\kappa$-Minkowski space \cite{MAJID1994348} etc. All of these models have various distinct properties, but the underlying idea and major consequences remain identical.

The Moyal spacetime was one of the initial models for NC spacetime, and it has been studied extensively covering various aspects like gauge theory in this space \cite{Wallet_2008}, study of gravity, spacetime curvature \cite{Harikumar2006} etc. One of the drawbacks of Moyal space is that it breaks Lorentz symmetry but this can be retained via a Hopf algebra approach \cite{LorentzinvarianceviaHopfalgebra}. In this paper, we will look at $\kappa$-Minkowski spacetime \cite{Lukierski1992dt} and some new unique features associated with it. $\kappa$-Minkowski spacetime is a type of noncommutative spacetime which has a Lie-algebra type commutation relation between its spacetime coordinates $\hat{x}_\mu$ i.e.,
\begin{equation}
    [\hat{x}_\mu,\hat{x}_\nu]=i C^\lambda_{\mu\nu}\hat{x}_\lambda
\end{equation}
and in our particular model of the $\kappa$-Minkowski space, this commutation relation simplifies to,
\begin{align}
    [\hat{x}_0,\hat{x}_i] &=ia\hat{x}_i \\
    [\hat{x}_i,\hat{x}_j] &=0
\end{align}
using this relation and various ansatz \cite{Meljanac2006} one can establish a $\kappa$-Poincaré algebra which mathematically is a Hopf algebra with a deformed coalgebra sector but with undeformed algebra sector (that is, the algebra is the standard Poincaré algebra). Here ansatz means various generic forms for the noncommutative quantities in terms of the corresponding commutative quantities and derivatives; we refer to this as a `realization' of noncommutative quantities. In this article, we will look at some properties of a Hamiltonian system in $\kappa$-Minkowski space. In such a space up to first order in the deformation parameter '$\alpha$' we have the modified Hamiltonian (in 1-D) \cite{Verma2018} as
\begin{equation}
    H_\alpha(x,p)=\frac{p^2}{2m_{eff}}+V(x+\alpha p)
\end{equation}
where $\alpha=a/2mc$ and $m_{eff}=m/(1+2m^2c^2\alpha)$(or in SI units $m_{eff}=m/(1+2m^2c^2\alpha/\hbar)$). We will also obtain some new features that arise because of the non trivial Hamiltonian which has a $p^4$ term. We will then study the tuning parameter of the system (which we will define later), entropy, and its derivatives as a function of the system's average energy, and find the critical energy numerically. We will also obtain new upper bounds on the noncommutative parameter by arguing that these transitions occur only at the Planck scale. The driving motivation behind this paper is the idea that introducing noncommutativity into spacetime creates structural changes and this should manifest not just as correction terms in standard physical quantities like fields, forces, etc. but at a more observational level. To demonstrate this, we study the $\kappa$-deformed lattice $\phi^4$ model. We will show that distinct behavior is seen in such systems that does not  vanish smoothly in the limit as $\alpha \rightarrow 0$ to give the standard results expected, that is, the system undergoes a discontinuous change in this property as $\alpha$ goes from being zero to a non zero value. In our case, we found that if $\alpha \ne0$ the particles in various lattice sites can occupy two degenerate momentum states even when the system is in the ground state, we argue that this is due to the momentum space topology being deformed. We then answer a possible follow up question whether this change can be observed at low energies. To answer this question, we derived a power law connecting the energy at which this transition occurs to the noncommutative parameter. There we have argued that the decay behavior of wave packets is affected by the noncommutativity of spacetime \cite{a2026wavepacketsasymptoticskappaminkowski}.

One of the powerful tools we will use repeatedly is singularity theory. A mathematically interested reader is referred to Arnold's \cite{Arnold1998} work for details. We also give a short introduction to the needed aspects in our paper on wave packet asymptotics. In section.~\ref{sec2} we introduce the topological theory of phase transitions. Then in section.~\ref{sec3} we introduce our model of study which is the $\kappa$-deformed 1-D lattice $\phi^4$ model which we analyze via the mean field approximation. In section.~\ref{sec4} we report the numerical results showing phase transitions of the model. In section.~\ref{sec5} we obtain the same conclusion of our numerical results by using arguments from singularity theory. In section.~\ref{sec 6} we give some numerical estimates of the energy scale at which one can see such topological phase transitions in momentum space. Finally in section.~\ref{sec6} we give some concluding remarks.

\section{Topological theory of phase transition}\label{sec2}
 The rationalization of Hamiltonian dynamics using techniques from Riemannian geometry and exploration of phase transitions from the point of view of topology has led to a topological theory of phase transition \cite{Gori2022}. This theory proves that a phase transition is necessarily followed by a change in the topology of the level manifold. Currently, two theorems have been formulated, and various models have been tested for this theorem. The theory only proves the correspondence in one direction, that is, if a phase transition occurs, then there is an associated change in the topology of the level manifold, but the converse is yet to be proven clearly although for most of the systems the converse has been shown to be true. However, there is one simple example that seems to contradict this theory, the lattice $\phi^4$ model. To explain this anomaly, arguments using asymptotic diffeomorphicity \cite{Gori2022} of the level sets were presented and the theory was modified accordingly. One of the driving results of this theory is the fact that the configurational entropy of the system is directly related to the Morse index of the critical points in the level manifold thus establishing a link between topology of the level manifold and the thermodynamics of the system, this is given by 
\begin{eqnarray}
S_N(v) &=& (k_B/N) \log \left[ \int_{M_v^{V_N}} d^N q \right]\notag \\ &=& \frac{k_B}{N} \log \left\{ vol \left[ M_v^{V_N} \setminus \bigcup_{i=1}^{\mathcal{N}(v)} \Gamma(x_c^{(i)}) \right] \right.\notag \\
&& \left. + \sum_{i=0}^N w_i \mu_i(M_v^{V_N}) + \mathcal{R} \right\} .
\end{eqnarray}
This means that any change in the Morse index of the system may be reflected as a thermodynamic phase transition. For a detailed account of this theory, we refer to a book by Pettini \cite{Pettini2007}. In the next section we will study the topological and thermodynamic properties of the $\kappa$-deformed 1-D mean field lattice $\phi^4$ model whose undeformed theory has been given in this paper \cite{PhysRevE.70.041101}. We follow the methods in the same paper and reproduce the numerical results, but for the $\kappa$-deformed model, and find the critical energy at which the transition occurs. Later we will justify our numerical result using singularity theory.

\section{$\kappa$-deformed 1-D mean field lattice $\phi^4$ model}\label{sec3}
This model has been extensively studied in the context of the topological theory of phase transitions in this paper \cite{PhysRevE.70.041101}, the standard undeformed Hamiltonian of this model \cite{PhysRevD.60.045006,PhysRevE.70.041101} is:
\begin{align} 
H &= \sum_{i} h(\phi_i) - \frac{J}{2N}\left(\sum_{i} \phi_i\right)^2 \\&= \sum_{i} h(\phi_i) - \frac{JNM^2}{2} , \\
h(\phi) &= -\frac{\phi^2}{2} + \frac{\phi^4}{4} ,\label{29}
\end{align}
Where $M=\sum_i\phi_i/N$ is the mean field approximation and $J$ is the coupling constant for the interaction between two lattice sites. In general when studying such system one is only concerned with the potential energy term of the Hamiltonian; here however, since we are interested in the deformed structure in which potential gets momentum dependent terms we must include the kinetic energy term also. 
\begin{align} 
H &= \sum_i \frac{p_i^2}{2m}+\sum_{i} h(\phi_i) - \frac{J}{2N}\left(\sum_{i} \phi_i\right)^2 \\&= \sum_i \frac{p_i^2}{2m}+\sum_{i} h(\phi_i) - \frac{JNM^2}{2} , \\
h(\phi) &= -\frac{\phi^2}{2} + \frac{\phi^4}{4} ,\label{30}
\end{align}
Now we apply the $\kappa$-deformation to this to get the $\kappa$- deformed Hamiltonian as 
\begin{eqnarray}
    H_{\alpha}(p,\phi) &=& \sum_i \frac{p_i^2}{2m_{eff}} + \sum_{i} h(\phi_i+\alpha p_i) 
    \nonumber \\ && - \frac{J}{2N}\left(\sum_{i} \phi_i+\alpha p_i\right)^2
\end{eqnarray}
In this article instead of looking at stationary points by taking derivatives with $\phi_j$ we will do it with respect to $p_j$. This is because the stationary points are no longer the trivial $p_i=0$ because of the deformed Hamiltonian. In this respect we will treat the variables $\phi_i$ as constant and without loss of generality keep them zero to just study the momentum space topology of the system which gets interesting new properties due to the $\kappa$-deformation. As a result, the mean field approach will be applied to define 
\begin{equation}
    l=\frac{1}{N} \left(\sum_i \phi_i+\alpha p_i\right)
\end{equation}
Also another important point to be careful about is the fact that when we defined $h(\phi)$ for the commutative case we defined it such that it has a double well structure now for $h(\phi_i+\alpha p_i)$ the same value for the coefficients will not work, so we will define a general coefficient and later fix it so that Hamiltonian will have a double well structure.
That is, if we take $h(\phi_i+\alpha p_i)=b(\phi_i+\alpha p_i)^2+(\phi_i+\alpha p_i)^4/4$  we then find that the condition for stationary points is 
\begin{equation}
    \frac{\partial H}{\partial p_j}=\left(\frac{1}{m_{eff}}+2b\alpha^2\right)p_j+\alpha^4 p_j^3 -Jl\alpha=0
\end{equation}
now we choose $\dfrac{1}{m_{eff}}+2b\alpha^2=-2$ (Note that the right-hand side has units of $1/M$, which we omit in subsequent steps to avoid notational complications). This implies that the stationary point condition is
\begin{equation}
    p_j^3-\frac{2}{\alpha^4}p_j-\lambda_1=0, \ \ \ \lambda_1=\frac{J l}{\alpha^3}
\end{equation}
Later, to solve the equation numerically, we will use a change of variables $p=u/\alpha^2,\lambda_{sc}=\lambda_1\alpha^6$ to scale the roots so that it also works for very small $\alpha$. This scaling allows us to obtain the same structure of equations as in the paper \cite{PhysRevE.70.041101} but in terms of the new variable $u=p\alpha^2$, but to directly see the effects of $\kappa$- deformation we will keep $\alpha$ in the picture.

Now we introduce our first concept from singularity theory; it is known that the number of real solutions of Eq.~15 can be found from the level bifurcation set of the polynomial. The level bifurcation set of a cubic polynomial $x^3+a x+b$ is given by $27b^2+4a^3=0$ \cite{Arnold1998} now substituting the parameters into this we obtain a critical value for $\lambda_1$ which is $\lambda_c=4\sqrt{2}/3\sqrt{3}\alpha^6$ so now we can analyze the situation when $|\lambda_1|<\lambda_c$, $|\lambda_1|=\lambda_c$ and $|\lambda_1|>\lambda_c$.
\subsection{Case 1: $|\lambda_1|<\lambda_c$}
In this case, there will be three distinct solutions, say $p_1,p_2,p_3$. Now each $p_i$ for $1\leq i\leq N$ will be one of these three solutions, thus we define the fraction of solutions for each of these three distinct solutions as $n_1,n_2,n_3$ now we can see that we have a system of equations
\begin{align}
    \sum_i n_i &=1\\
    \lambda_1\alpha^3 &=Jl=J\sum_i n_i\alpha p_i\\
    v &=\frac{H_\alpha(p_i)}{N} =\sum_i (n_i \frac{p_i^2}{2m_{eff}}+n_ih(\alpha p_i))-\frac{\lambda_1^2\alpha^6}{2J}
\end{align}
Now, by choosing various values of $\lambda_1$ and $v$, the system can be solved for the fractions $n_1,n_2,n_3$. This is what we will do later numerically, and once we find the fractions, we will define the configuration entropy of the system as a Shannon-type entropy, that is, $S=-\sum_i n_i \log{n_i}$. Then we will maximize this with respect to $\lambda_1$ and express it as a function of the average energy $v$ and then by plotting this we will be able to identify phase transitions. If one wants to see how the topology of the momentum space changes as $|\lambda_1|$ passes through the critical value, one can evaluate the Morse index, which is the number of negative eigenvalues of the Hessian at the critical point, and we will see that it is indeed different, but we will, however, not do this calculation here.
\subsection{Case 2: $|\lambda_1|=\lambda_c$}
In this case there are only two real solutions, say $p_1,p_2$ and so we have only two fractions of the number of particles in these stationary states, say $n_1,n_2$ and hence we only need the first two equations from above. In this case the solutions for the cubic equation can be explicitly found and are given by $p_1=2\sqrt{2}/\alpha ^2\sqrt{3}$ and $p_2=-\sqrt{2}/\alpha^2\sqrt{3}$ now we can solve Eq.~16 and Eq.~17 since every parameter is known. We get as solutions
\begin{align}
    n_1 &=\frac{1}{3}\left(1 + \dfrac{l\alpha}{\sqrt{6}}\right)\\
    n_2 &=\frac{2}{3} \left(1 - \dfrac{l\alpha}{\sqrt{6}} \right) 
\end{align}
The most interesting fact about this result is that these fractions in the limit as $\alpha$ goes to zero converge to $1/3$ and $2/3$ so this result persists no matter how small $\alpha$ is. In the commutative case there are no multiple momentum values available for the system since the Hamiltonian has $A_1$ singularity instead of $A_3$  (as in noncommutative case). This means the fraction $n_1$ is always one. If we compares these two cases we see that this behavior is discontinuous in $\alpha$ since $\alpha=0$ (commutative case) and arbitrary small values of $\alpha$ gives different values for the fractions.

Typically adding noncommutativity into the picture is expected to give perturbative correction to standard quantities making it exceptionally hard to detect experimentally but what we propose here is that if the spacetime is in fact noncommutative the effects will be seen at a more structural or topological level rather than just corrections. Because, as we can see, the commutative case gives the trivial $p_i=0$ stationary points, and then these fractions cannot be solved uniquely from Eq.~16, Eq.~17.

In short what we are proposing here is that as we tune the parameter $\lambda_1$ by changing $J$ so that it reaches $\lambda_c$ we see a momentum space distribution where about one third of particles occupy one stationary state and the other two thirds occupies the other stationary state. This result is robust and insensitive to the actual value of $\alpha$, however, the one problem we might face to see this result in the lab will be that the critical value $\lambda_c$ will be so high (since it is inversely proportional to $\alpha^6$) that we will need a very large value of $J$ to reach this point, but this is expected as we are trying to probe the noncommutative properties of spacetime. 
\subsection{Case 3: $|\lambda_1|>\lambda_c$}
In this case there exists only one real solution say $p_1$ and so we only have to solve for one fraction, then it is trivial that this fraction is one, as all particles will be in this unique stationary state, hence from Eq.~17 we get
\begin{equation}
    \lambda_1=\frac{1}{\alpha^2}Jp_1
\end{equation}
Now we can substitute this into Eq.~15 to get
\begin{equation}
    p_1^3-\frac{2}{\alpha^4}p_1=\frac{1}{\alpha^2}Jp_1
\end{equation}
but since $|\lambda_1|>\lambda_c>0$, $p_j=0$ will not be a solution of Eq. (7) so the only possible solutions of Eq. (14) are 
\begin{equation}
    p_1=\pm \sqrt{\frac{J}{\alpha^2}+\frac{2}{\alpha^4}}
\end{equation}
but if we have $J>0$ then Eq. (7) cannot have the negative solution from above since $\lambda_1>0$ so the only solution is the positive solution from above.

Next we will look at the numerical solution of Eqs. (8), (9), and (10), for doing this, we will plot the solutions $n_1,n_2,n_3$ and entropy as functions of $v$. We will maximize the entropy with respect to $\lambda_1$ as a result the maximum contribution will happen from case 1 where $|\lambda_1|<\lambda_c$ where we have more available stationary states and compared to this the contribution from the other two cases will be small. After plotting, we will see that varying $J$ induces distinct changes in the curves indicating new phase transitions; finally, we will vary $\alpha$ and see how the critical value of energy $v_c$ depends on the noncommutative parameter $\alpha$.
\section{numerical results}\label{sec4}
We solve Eqs. (8), (9), and (10) by performing appropriate scaling so that the numerical results are not affected due to very low values of $\alpha$, and the results obtained are summarized below

As we can observe, the smoothness of the curves depends on the values of $J$ and $\alpha$.\\
Finally, we look at the relation between $\alpha$ and $v_c$ (which we call $\bar{E}$, which is the value of $v$ at which the curve of $\lambda_1$ becomes non differentiable).
From the Log Log plot of $\bar{E}$ versus $1/\alpha$ it is clear that there is a power relation between the two quantities which we can find from the graph and it turns out that 
\begin{equation}
    \bar{E}\sim \frac{1}{\alpha^4}
\end{equation}
This is one of the main results of our paper, once the system average energy reaches the order of $\alpha^{-4}$ a topological phase transition occurs in momentum space, which enables particles to access multiple degenerate momentum states even when they are at the minima of their potential energy. What is even more remarkable is the fact that this behavior can be predicted via singularity theory without any numerical calculations. We do this by using the swallow tail structure of the level bifurcation set of the Hamiltonian.
\section{Explanation via singularity theory}\label{sec5}
We define the single-particle Hamiltonian at $x=0$. This isolates our study entirely to the momentum sector by evaluating the system exactly at the stationary point of the undeformed potential $V(x)$. Consequently, we are specifically analyzing how the momentum space topology behaves at the spatial minimum, yielding the Hamiltonian: 
\begin{equation}
    H_1(p)=\frac{p^2}{2m_{eff}}+h(\alpha p)\label{eq 25}
\end{equation}
Where $h(\alpha p)=b\alpha^2p^2+\alpha^4p^4/4$
with the same choice for the parameter $b$ now we look at a polynomial that is nothing but the level surface given by 
\begin{equation}
    H_1(p)=\frac{p^2}{2m_{eff}}+h(\alpha p) =E
\end{equation}
This leads to
\begin{equation}
    -p^2+\alpha^4 \frac{p^4}{4}=E
\end{equation}
which can be written as
\begin{equation}
    p^4 -\frac{4}{\alpha^4}p^2-\frac{4}{\alpha^4}E=0\label{eq 28}
\end{equation}
This is a quartic polynomial in $p$ whose level bifurcation set is given by the swallowtail structure. For a general quartic polynomial in $p$ its miniversal deformation is $p^4+\lambda_1 p^2+\lambda_2p+\lambda_3$ its level bifurcation set consists of those values of $\lambda_i$ that make the polynomial have multiple roots, which is of the form $(p-a)^2(p-b)(p-c)$, so we have the equation
\begin{equation}
    (p-a)^2(p-b)(p-c)=p^4+\lambda_1 p^2+\lambda_2p+\lambda_3
\end{equation}
This gives us a system of equations
\begin{align}
    -3a^2+bc &=\lambda_1\\
    2a^3-2abc &=\lambda_2\\
    a^2bc&=\lambda_3
\end{align}
Where $\lambda_1=-4/\alpha^4$,$\lambda_2=0$ and $\lambda_3=-4E/\alpha^4$
This gives an algebraic equation in $\lambda_i$ that represents the swallowtail structure. We now find the cross section of this structure when $\lambda_2=0$ since our polynomial does not have a linear term, using this fact the system of equations gives
\begin{figure*}[htb]
    \centering
    \includegraphics[width=0.8\textwidth]{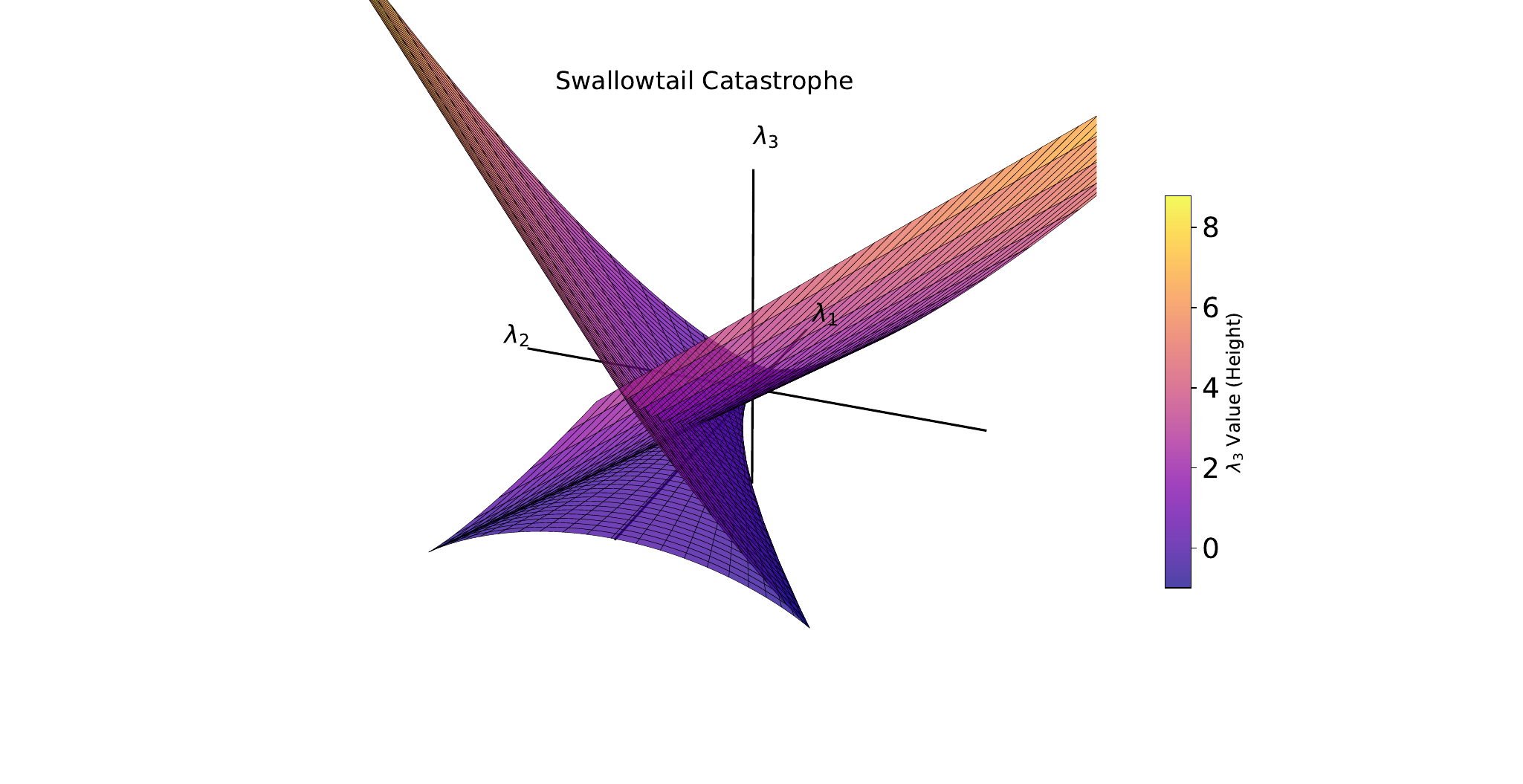}
    \caption{The swallowtail catastrophe is plotted in the figure, this is the level bifurcation set of the miniversal deformation of $A_3$ singularity $p^4+\lambda_1p^2+\lambda_2p+\lambda_3$}
    \label{fig:placeholder1}
\end{figure*}
\begin{align}
    \lambda_3 &=\frac{\lambda_1^2}{4}\\
    \implies E_c &\sim-\frac{1}{\alpha^4}\\
    \implies |E_c| &\sim\frac{1}{\alpha^4}
\end{align}
This justifies the power law dependence of the critical value of the average energy as seen from the Log-Log plot in Fig.4\footnote{One has to be careful about the units here. The numerical data shown on the plots are from the calculation treating the paramters like $x$,$p$,$E$ as dimensionless parameters by scaling them appropriately by $\hbar/mc$,$mc$ and $mc^2$ respectively. Any further calculation in SI or other units must be done after accounting for this.}.
\section{A numerical estimate of the Critical Energy}\label{sec 6}
Before we see some numerical estimate of the critical energy we must restore all units and make the equations match dimensionally. Starting from Eq.~\ref{eq 25} we get
\begin{align}
    \frac{p^2}{2m_{eff}}+b\alpha^2p^2+\frac{1}{4}\alpha^4p^4&=E\\
    \implies (\frac{1}{2m_{eff}}+b\alpha^2)p^2+\frac{1}{4}\alpha^4p^4&=E
\end{align}
to make this dimensionally consistent we introduce two parameters $q$ and $g$ with numerical value one and dimensions of $[M^{-1}]$ and $[ML^{-2}T^{-2}]$. Therefore we have 
\begin{align}
    -qp^2+\frac{1}{4}g\alpha^4p^4=E
\end{align}
Now we can safely estimate the energy order in joules.
To see an estimate of this energy, we use some known bounds on $a$. From experimental violation of the equivalence principle one gets the following.
\begin{align}
   \frac{|a|mc}{\hbar} &\leq 10^{-13}\\
    \implies |\alpha| &=|\frac{a}{2mc}|=\frac{\hbar}{2m^2 c^2}|\frac{amc}{\hbar}|\\
    &\leq \frac{\hbar}{2m^2 c^2}\times10^{-13} \approx 7.08\times10^{-5}
\end{align}
This gives an estimate of the order of energy at $10^{20}$ Joules. Which is several orders higher than the Planck scale energy.
%where we have used electron mass for $m$, this gives us a lower bound for critical energy of the order of $10^4 J$ which is clearly achievable in terms of energy scales that are used today, now if we insist that such transition exists only at very high energy, then from this we can get a better bound for the noncommutative  parameter. Say, the transition occurs at nearly the Planck energy which is of order $10^9$ joules which gives an $\alpha\approx5.6\times10^{-3}$ which gives an upper bound for $a$ in the order of $10^{-26}$ joules, whereas if we look only at Eq.29 we get an upper bound for $a$ in the order of $10^{-24}$ joules, so if we find that such topological phase transitions occur at an energy comparable to the Planck scale then the noncommutative parameter is at least a hundred times smaller than we thought.
\section{conclusion}\label{sec6}
Various models of quantum gravity have come forward in recent years, and various theoretical advancements have occurred in this field, but experimental verification for these models has been minimal. This is often due to the high energy required to probe the quantum effects of spacetime, which is of the order of Planck energy. Noncommutative spacetimes have been used in various models of quantum gravity like string theory \cite{NathanSeiberg_1999}, doubly special relativity, LQG etc., and therefore any evidence of a noncommutative spacetime will greatly affect the advancement of the field of quantum gravity. Due to extremely tiny correction effects of noncommutative spacetimes it is nearly impossible to attribute such observations to the possibility of a noncommutative spacetime. As a result, one must look at much more drastic changes that happen in the behavior of systems once we assume that spacetime is noncommutative. We looked at one such change that is not simply a correction term in a physical quantity. It is such results that we must use to convince ourselves that spacetime might indeed be noncommutative. The only problem is that to see such changes one might have to reach large energy scales often of the order of Planck energy. From the power law relation, once we get the critical energy we can find an exact estimate for the noncommutative parameter from the value of $\alpha$; this opens up various experimental and theoretical means to prove the existence of a noncommutative spacetime.
\section{Acknowledgements}
We acknowledge the support of the NIUS programme of HBCSE-TIFR funded by the Department of Atomic
Energy, Govt. of India (Project No. RTI4001)
\begin{figure*}[t]
    \centering
    \includegraphics[page=1, width=0.85\linewidth]{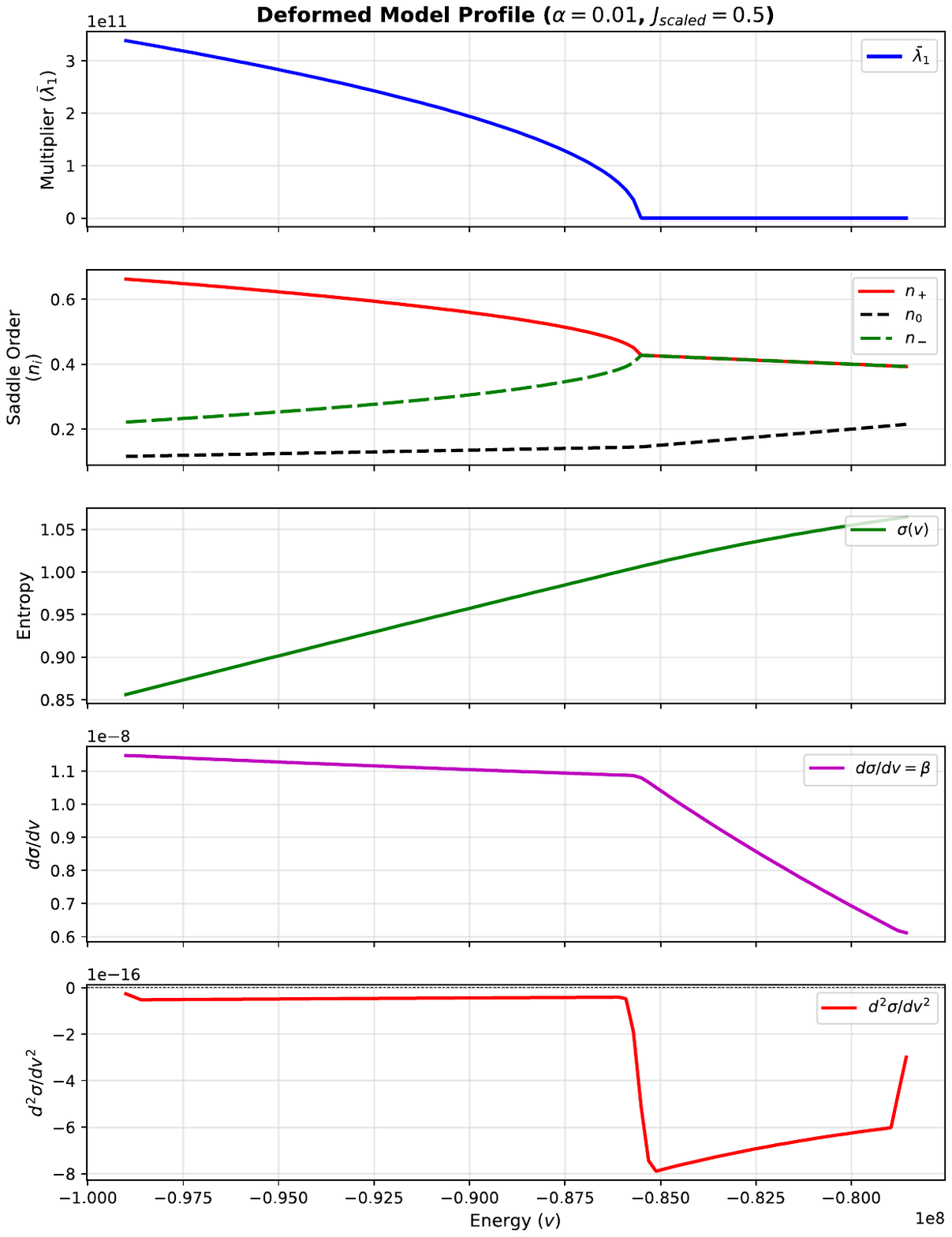}
    \caption{In this figure we have plotted the tuning parameter, occupancy fractions, entropy and its first two derivatives we observe a sharp point in the tuning parameter $\lambda_1$ corresponding to which we observe sharp changes in the first and second derivative of entropy hinting at a phase transition}
    \label{fig:placeholder2}
\end{figure*}
\begin{figure*}[t]
    \centering
    \includegraphics[page=2, width=0.85\linewidth]{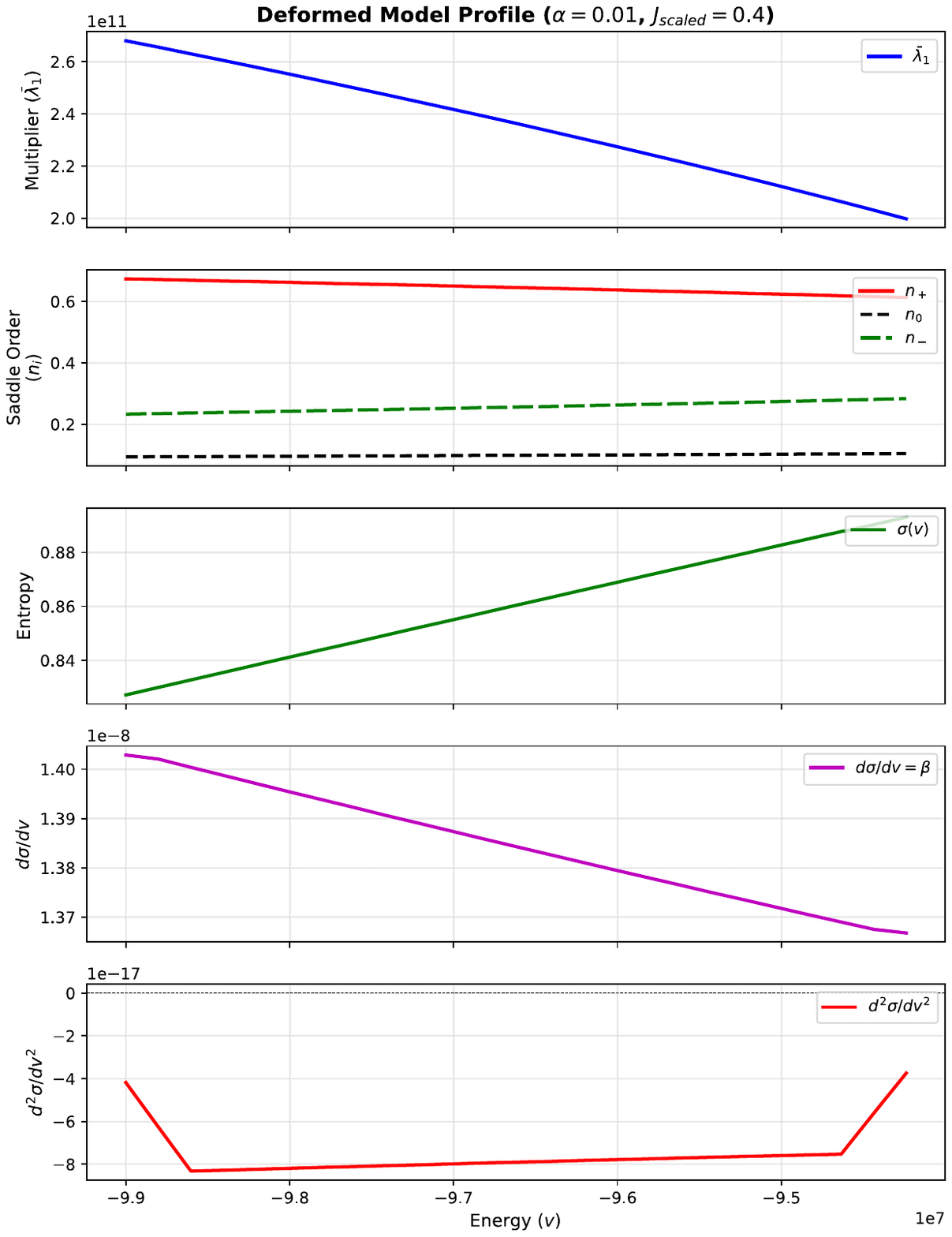}
    \caption{here as we change the values of $J_{scaled}$ we see changes in the behavior of the curves namely that there is no sharp change in the tuning parameter}
    \label{fig:placeholder3}
\end{figure*}
\begin{figure*}[t]
    \centering
    \includegraphics[page=3, width=0.85\linewidth]{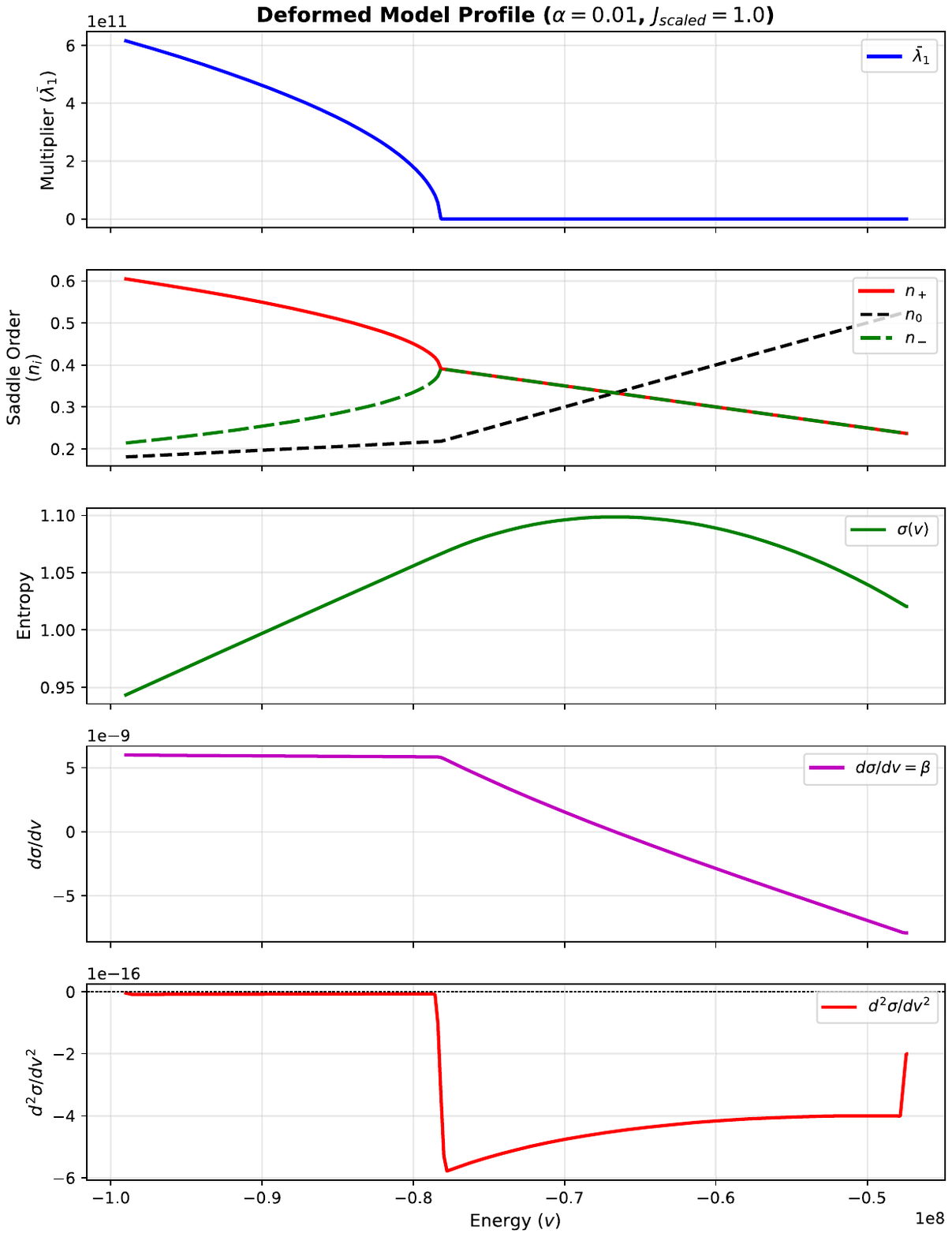}
    \caption{We keep the value of $\alpha$ fixed and change the value of $j_{scaled}$ to study its effect on phase transition.} 
    \label{fig:placeholder4}
\end{figure*}
\begin{figure*}[t]
    \centering
    \includegraphics[page=4, width=0.85\linewidth]{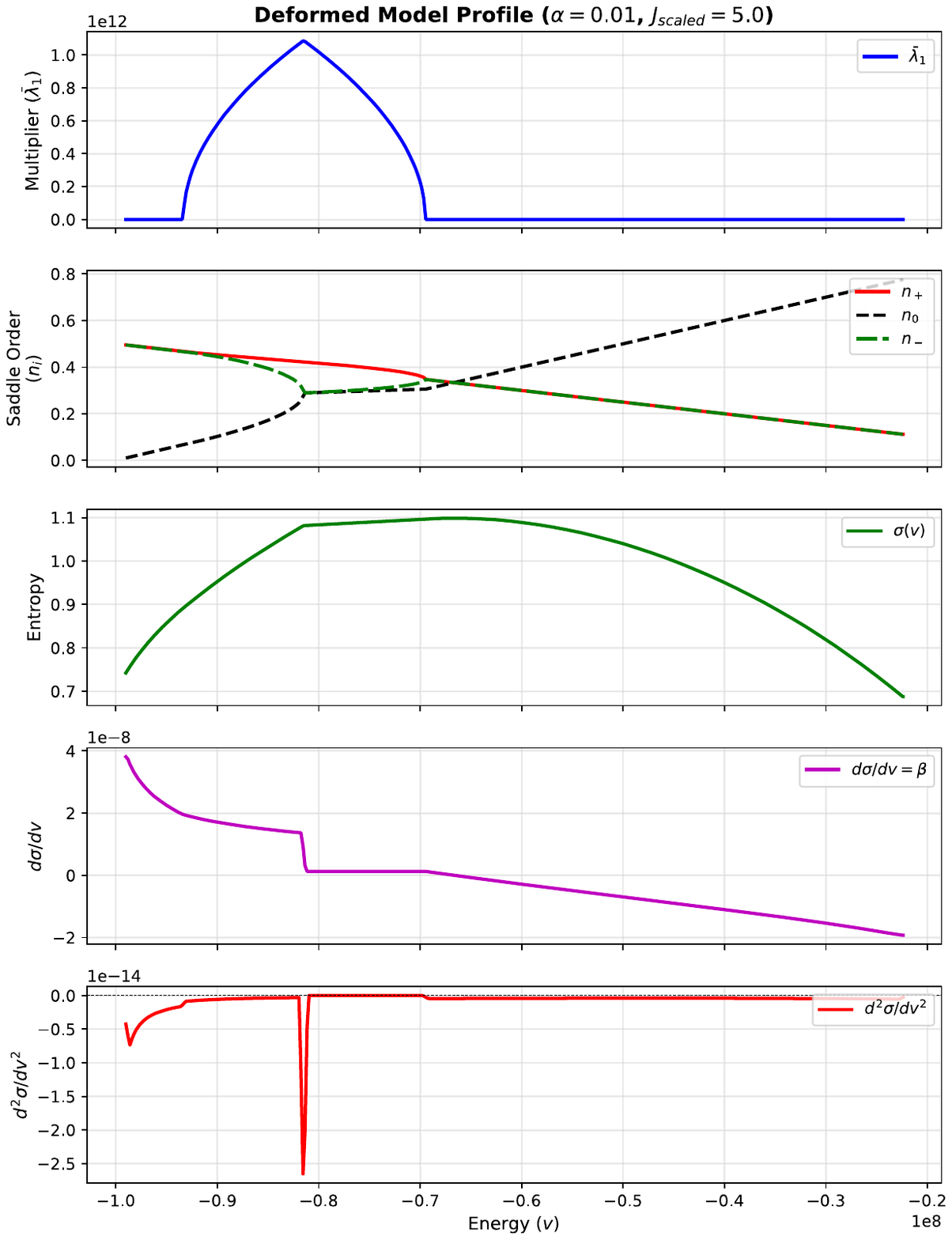}
    \caption{We see that for a fixed $\alpha$ increasing $J_{scaled}$ creates more critical points in the second derivative of entropy hinting at multiple phase transitions.} 
    \label{fig:placeholder5}
\end{figure*}
\begin{figure*}
    \centering
    \includegraphics[page=5, width=0.85\linewidth]{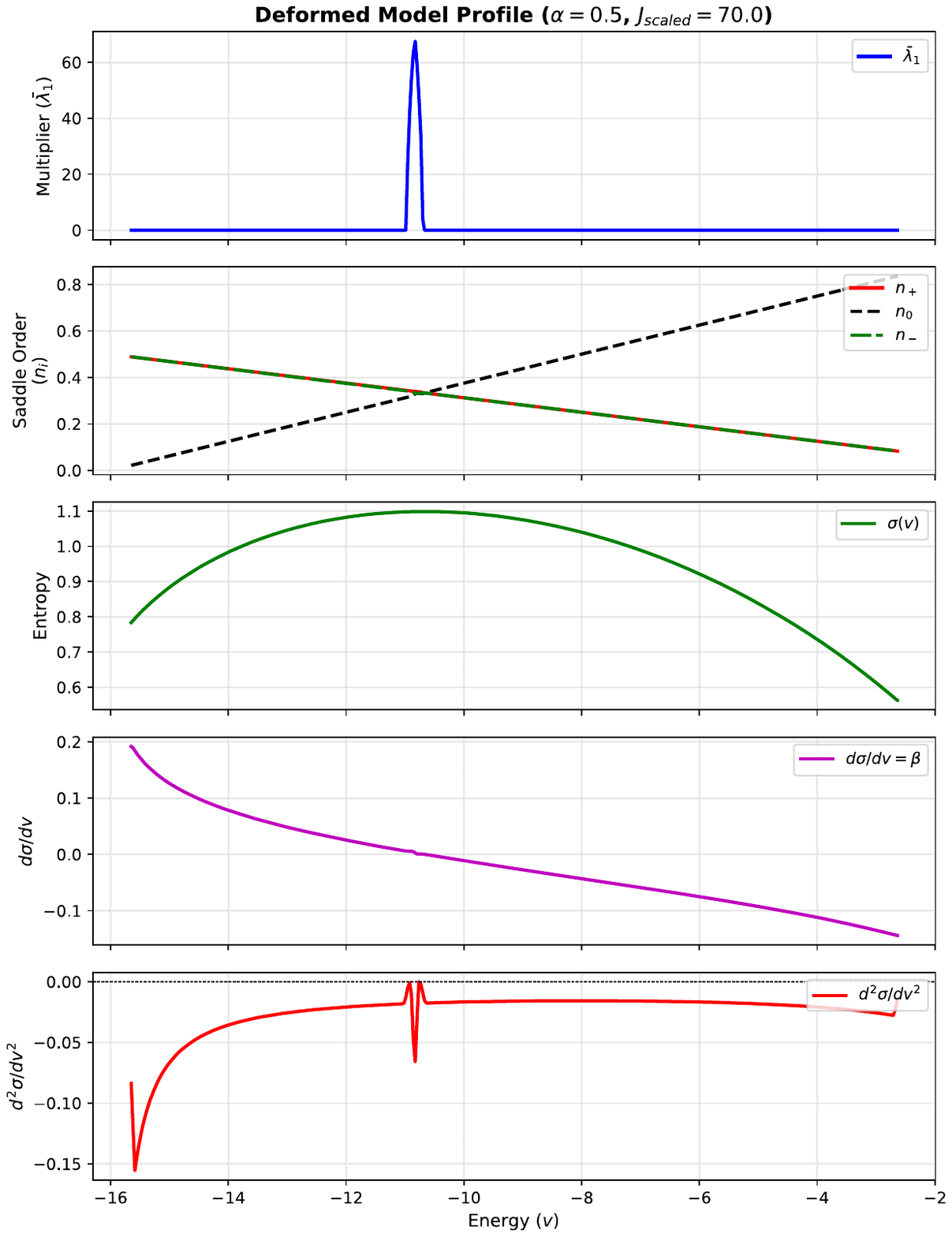}
    \caption{We see that the number of critical points is also sensitive to the value of $\alpha$.} 
    \label{fig:placeholder6}
\end{figure*}

\begin{figure*}
    \centering
    \includegraphics[page=7,angle=-90,trim=0 0 2.5in 0, clip, width=0.85\linewidth]{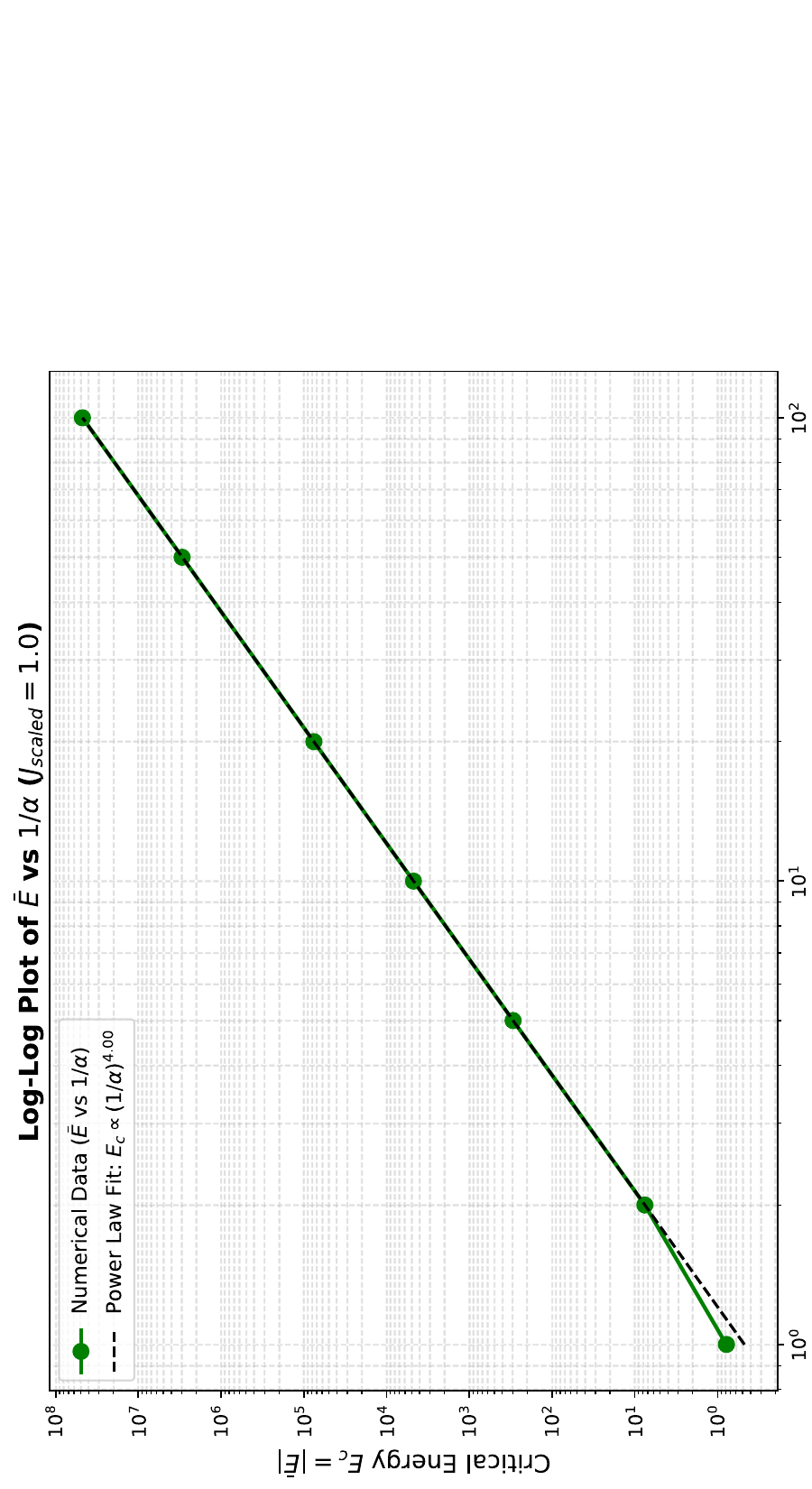}
    \caption{The figure shows the plot between the critical value of average energy (which is the value of $v$ at which the curve of tuning parameter $\lambda_1$ first becomes non differentiable when we vary $v$ from $0$ to negative that is the first sharp point in the curve of tuning parameter) and  $1/\alpha$ the plot shows a power law dependence of $E_c$ on $1/\alpha$ that is $E_c=1/\alpha^4$} 
    \label{fig:placeholder7}
\end{figure*}
\bibliographystyle{apsrev4-2}
\bibliography{ref.bib}

@article{RevModPhys.73.977,
  title = {Noncommutative field theory},
  author = {Douglas, Michael R. and Nekrasov, Nikita A.},
  journal = {Rev. Mod. Phys.},
  volume = {73},
  issue = {4},
  pages = {977--1029},
  numpages = {0},
  year = {2001},
  month = {Nov},
  publisher = {American Physical Society},
  doi = {10.1103/RevModPhys.73.977},
  url = {https://link.aps.org/doi/10.1103/RevModPhys.73.977}
}

@article{NathanSeiberg_1999,
doi = {10.1088/1126-6708/1999/09/032},
url = {https://doi.org/10.1088/1126-6708/1999/09/032},
year = {1999},
month = {oct},
publisher = {},
volume = {1999},
number = {09},
pages = {032},
author = {Nathan Seiberg and Edward Witten},
title = {String theory and noncommutative geometry},
journal = {Journal of High Energy Physics}
}

@book{connes1995noncommutative,
  title={Noncommutative Geometry},
  author={Connes, A.},
  isbn={9780080571751},
  lccn={94026550},
  url={https://books.google.co.in/books?id=QofCkAJjreYC},
  year={1995},
  publisher={Academic Press}
}

@article{Wallet_2008,
doi = {10.1088/1742-6596/103/1/012007},
url = {https://doi.org/10.1088/1742-6596/103/1/012007},
year = {2008},
month = {feb},
publisher = {},
volume = {103},
number = {1},
pages = {012007},
author = {J-C Wallet},
title = {Noncommutative induced gauge theories on Moyal spaces},
journal = {Journal of Physics: Conference Series}
}

@article{LorentzinvarianceviaHopfalgebra,
author = {Chaichian, Masud and Kulish, P.P. and Nishijima, K. and Tureanu, A.},
year = {2004},
month = {08},
pages = {98-102},
title = {On a Lorentz-Invariant Interpretation of Noncommutative Space-Time and Its Implications on Noncommutative QFT},
volume = {604},
journal = {Physics Letters B},
doi = {10.1016/j.physletb.2004.10.045}
}

@article{Lukierski1992dt,
    author = "Lukierski, Jerzy and Nowicki, Anatol and Ruegg, Henri",
    title = "{New quantum Poincare algebra and k deformed field theory}",
    reportNumber = "UGVA-DPT-1992-07-776, LPTB-92-04",
    doi = "10.1016/0370-2693(92)90894-A",
    journal = "Phys. Lett. B",
    volume = "293",
    pages = "344--352",
    year = "1992"
}

@article{Meljanac2006,
  author    = {Meljanac, S. and Stoji{\'{c}}, M.},
  title     = {New realizations of Lie algebra kappa-deformed Euclidean space},
  journal   = {The European Physical Journal C - Particles and Fields},
  volume    = {47},
  number    = {2},
  pages     = {531--539},
  year      = {2006},
  month     = {Aug},
  doi       = {10.1140/epjc/s2006-02584-8},
  url       = {https://doi.org/10.1140/epjc/s2006-02584-8},
  issn      = {1434-6052}
}

@article{Verma2018,
doi = {10.1209/0295-5075/122/40001},
url = {https://doi.org/10.1209/0295-5075/122/40001},
year = {2018},
month = {jun},
publisher = {EDP Sciences, IOP Publishing and Società Italiana di Fisica},
volume = {122},
number = {4},
pages = {40001},
author = {Verma, Ravikant and Ghorai, Debabrata and Gangopadhyay, Sunandan},
title = {Path integral action of a particle in κ-Minkowski spacetime},
journal = {Europhysics Letters}
}

@Inbook{Arnold1998,
author="Arnold, V. I.
and Goryunov, V. V.
and Lyashko, O. V.
and Vasil'ev, V. A.",
title="Critical Points of Functions",
bookTitle="Singularity Theory I",
year="1998",
publisher="Springer Berlin Heidelberg",
address="Berlin, Heidelberg",
pages="10--50",
isbn="978-3-642-58009-3",
doi="10.1007/978-3-642-58009-3_1",
url="https://doi.org/10.1007/978-3-642-58009-3_1"
}

@article{Gori2022,
doi = {10.1088/1751-8121/ac7f09},
url = {https://doi.org/10.1088/1751-8121/ac7f09},
year = {2022},
month = {aug},
publisher = {IOP Publishing},
volume = {55},
number = {37},
pages = {375002},
author = {Gori, Matteo and Franzosi, Roberto and Pettini, Giulio and Pettini, Marco},
title = {Topological theory of phase transitions},
journal = {Journal of Physics A: Mathematical and Theoretical}
}

@Inbook{Pettini2007,
title="Topological Hypothesis on the Origin",
bookTitle="Geometry and Topology in Hamiltonian Dynamics and Statistical Mechanics",
year="2007",
publisher="Springer New York",
address="New York, NY",
pages="203--227",
isbn="978-0-387-49957-4",
doi="10.1007/978-0-387-49957-4_7",
url="https://doi.org/10.1007/978-0-387-49957-4_7"
}

@article{PhysRevE.70.041101,
  title = {Topological properties of the mean-field ${\ensuremath{\phi}}^{4}$ model},
  author = {Andronico, A. and Angelani, L. and Ruocco, G. and Zamponi, F.},
  journal = {Phys. Rev. E},
  volume = {70},
  issue = {4},
  pages = {041101},
  numpages = {14},
  year = {2004},
  month = {Oct},
  publisher = {American Physical Society},
  doi = {10.1103/PhysRevE.70.041101},
  url = {https://link.aps.org/doi/10.1103/PhysRevE.70.041101}
}

@article{PhysRevD.60.045006,
  title = {Antiferromagnetic ${\ensuremath{\varphi}}^{4}$ model. I. The mean-field solution},
  author = {Branchina, Vincenzo and Mohrbach, Herv\`e and Polonyi, Janos},
  journal = {Phys. Rev. D},
  volume = {60},
  issue = {4},
  pages = {045006},
  numpages = {13},
  year = {1999},
  month = {Jul},
  publisher = {American Physical Society},
  doi = {10.1103/PhysRevD.60.045006},
  url = {https://link.aps.org/doi/10.1103/PhysRevD.60.045006}
}

@article{JACKIW200230,
title = {Physical instances of noncommuting coordinates},
journal = {Nuclear Physics B - Proceedings Supplements},
volume = {108},
pages = {30-36},
year = {2002},
note = {LIGHT-CONE PHYSICS: PARTICLES AND STRINGS Proceedings of the International Workshop TRENTO 2001},
issn = {0920-5632},
doi = {https://doi.org/10.1016/S0920-5632(02)01302-6},
url = {https://www.sciencedirect.com/science/article/pii/S0920563202013026},
author = {R. Jackiw}
}

@article{Synder,
  title = {Quantized Space-Time},
  author = {Snyder, Hartland S.},
  journal = {Phys. Rev.},
  volume = {71},
  issue = {1},
  pages = {38--41},
  numpages = {0},
  year = {1947},
  month = {Jan},
  publisher = {American Physical Society},
  doi = {10.1103/PhysRev.71.38},
  url = {https://link.aps.org/doi/10.1103/PhysRev.71.38}
}

@ARTICLE{fuzzysphere,
       author = {{Madore}, J.},
        title = "{The fuzzy sphere}",
      journal = {Classical and Quantum Gravity},
         year = 1992,
        month = jan,
       volume = {9},
       number = {1},
        pages = {69-87},
          doi = {10.1088/0264-9381/9/1/008},
       adsurl = {https://ui.adsabs.harvard.edu/abs/1992CQGra...9...69M}
}

@article{MAJID1994348,
title = {Bicrossproduct structure of κ-Poincare group and non-commutative geometry},
journal = {Physics Letters B},
volume = {334},
number = {3},
pages = {348-354},
year = {1994},
issn = {0370-2693},
doi = {https://doi.org/10.1016/0370-2693(94)90699-8},
url = {https://www.sciencedirect.com/science/article/pii/0370269394906998},
author = {S. Majid and H. Ruegg}
}

@article{WESS1991302,
title = {Covariant differential calculus on the quantum hyperplane},
journal = {Nuclear Physics B - Proceedings Supplements},
volume = {18},
number = {2},
pages = {302-312},
year = {1991},
issn = {0920-5632},
doi = {https://doi.org/10.1016/0920-5632(91)90143-3},
url = {https://www.sciencedirect.com/science/article/pii/0920563291901433},
author = {Julius Wess and Bruno Zumino}
}

@book{Connes2,
    author = {Connes, Alain and Marcolli, Matilde},
    title = {Noncommutative Geometry, Quantum Fields and Motives},
    publisher = {American Mathematical Society, Hindustan Book Agency} ,
    year = {2008} ,
    address = {New Delhi, India} ,
    edition = {1st} ,
    isbn = {978-1-4704-5045-8}
}

@book{khalkhali,
author = {Khalkhali, Masoud},
title = {Basic Noncommutative Geometry},
publisher = {European Mathematical Society}, 
year = {2013},
address = {Z\"urich, Switzerland},
edition = {2nd}, 
isbn = {978-3-03719-128-6}
}

@article{Harikumar2006,
doi = {10.1088/0264-9381/23/24/024},
url = {https://doi.org/10.1088/0264-9381/23/24/024},
year = {2006},
month = {nov},
publisher = {},
volume = {23},
number = {24},
pages = {7551},
author = {Harikumar, E and Rivelles, Victor O},
title = {Noncommutative gravity},
journal = {Classical and Quantum Gravity}
}

@book{PauliBriefwechselVol2,
  author    = {Pauli, Wolfgang},
  title     = {Wissenschaftlicher Briefwechsel mit Bohr, Einstein, Heisenberg u.a., Band II: 1930--1939},
  editor    = {von Meyenn, Karl},
  publisher = {Springer-Verlag},
  address   = {Berlin, Heidelberg},
  year      = {1985},
  doi       = {10.1007/978-3-642-88741-3}
}

@book{PauliBriefwechselVol3,
  author    = {Pauli, Wolfgang},
  title     = {Wissenschaftlicher Briefwechsel mit Bohr, Einstein, Heisenberg u.a., Band III: 1940--1949},
  editor    = {von Meyenn, Karl},
  publisher = {Springer-Verlag},
  address   = {Berlin, Heidelberg},
  year      = {1993},
  doi       = {10.1007/978-3-540-78802-7}
}

@misc{a2026wavepacketsasymptoticskappaminkowski,
      title={Wave packets asymptotics in $\kappa$-minkowski space}, 
      author={A. Adith  and A. Bhagwat},
      year={2026},
      eprint={2610.12331},
      archivePrefix={arXiv},
      primaryClass={math-ph},
      url={https://arxiv.org/abs/2610.12331}, 
}
\end{document}